# Philosophy and problems of the definition of Extraterrestrial Life[1]

Jean Schneider
LUTh – Paris Observatory
Jean.Schneider@obspm.fr
5, place Jules Janssen
+33 1 45 07 77 77

running title:
Philosophy of the definition of exo-life

**Abstract**

When we try to search for extraterrestrial life and intelligence, we have to follow some guidelines. The first step is to clarify what is to be meant by "Life" and "intelligencee", *i.e.* an attempt to define these words. The word ``definition'' refers to two different situations. First, it means an arbitrary convention. On the other hand it also often designates an attempt to clarify the content of a pre-existing word for which we have some spontaneous preconceptions, whatever their grounds, and to catch an (illusory) "essence" of what is defined. It is then made use of pre-existing plain language words which carry an *a priori* pre-scientific content likely to introduce some confusion in the reader's mind.

The complexity of the problem will be analyzed and we will show that some philosophical prejudice is unavoidable. There are two types of philosophy: "Natural Philosophy", seeking for some essence of things, and "Critical (or analytical) Philosophy", devoted to the analysis of the procedures by which we claim to construct a reality. An extension of Critical Philosophy, Epistemo-Analysis (i.e. the Psycho-Analysis of concepts) is presented and applied to the

---

1 Presented at the Conference « The History and Philosophy of Astrobiology » Tycho Brahe's island Ven, Sverige (27-28 September 2011)

defintion of Life and to Astrobiology.

**Introduction**

On Earth, Life is perceived under two aspects: organic life and psychical life. Organic life, subject of Biology, is shared by all livings, from bacteria to humans. Psychical life is the attribute of some animals and humans. In the generally shared common view, psychical life culminates in human « intelligence » and there is no rupture, no fundamental gap between human intelligence and animal psychology. Intelligence is then just viewed as a skill, an ability to react to situations and the environment. In the context of Astrobiology, the question naturally arises whether these approaches are adapted to exo-life. Here we treat these questions with a philosophical approach. It must be pointed out that there are two types of Philosophy of Knowledge: « Natural » Philosophy and Critical Philosophy. Hereafter we first clarify some differences between these two conceptions of Philosophy. We then explain why critical Philosophy is more efficient than Natural Philosophy. We finally make an analysis, based on critical Philosophy and its extension called « Epistemo-Analysis », to try to define organic and intelligent exo-life. For the latter we will point out its basic difficulty. We know that there is no form of evolved Life elsewhere in the Solar System. But there is plenty of room for evolved life on extrasolar planets. We therefore deal here only with life on these exoplanets. To conclude this introduction, we underline that our discussion is inspired by its pragmatic consequences: what actions to take to search for organic and psychical exo-Life?

**Natural versus Critical Philosophy**

The word « Philosophy » covers a wide continent, with unclear borders and regions, such as Political Philosophy, Ethics, and Philosophy of Konowledge. The latter can be divided into two categories, Natural Philosophy and Critical Philosophy.

*Natural Philosophy*

Natural philosophy is based on the belief that there exists a « Reality » and that Knowledge has to catch its essence in statements called « The Truth ». Knowledge then acts as some identification between the subject's mind and the intimate essence of nature, based on opinions and convictions rather than on critics and analysis, like in religious beliefs and faith. It results that there is a tendency of projection of human feelings on the external world, so that Natural Philosophy has a flavor of animism. This was already pointed out by Bachelard (2002) as the obstacle of "substancialism" and of animism in his book *The Formation of Scientific Mind*.

Note that believers in natural Philosophy are rather insensitive to the analysis between the two approaches of Philosophy, since they convictions are based on a kind of faith.

*Critical Philosophy*

Critical Philosophy starts with an analysis of the procedures by which we explain, thanks to natural language, our various experience, in any domain. This explanation is called a "theory", or more generally a discourse. This approach was thoroughly developed in Kant's *Critique of Pure Reason* and reassessed in the context of modern science by E. Cassirer (1965) in *The Philosophy of Symbolic Forms*. It has been remarkably summarized in the introduction of the *Critique*: « If our knowledge starts WITH experience, it does not prove that it only derives FROM experience, since it could well be that even our experience-based knowledge is a composite of what we receive from our perceptions and of what our power to know [*i.e.* concepts] produces itself ». More recently, various authors, following the so-called school of Analytical Philosophy, have pointed out that an unavoidable instrument to explain our experience is natural language. In this view, our knowledge is always a construction, with the help of language, of a so-called « reality » which does not pre-exist, and not the discovery of an essence of pre-existing things. The idea of a reality as source of perception is thus purely metaphysical and in this sense realism is an idealism.

Opposite to Natural Philosophy, the approach of critical Philosophy has always been fruitful in science: to give just an example, it has helped to get rid of the notion of aether in Physics.

To summarize, Critical Philosophy deals with the processes of construction of a "reality", while Natural Philosophy deals with an essence of things (illusory from the point of view of critical philosophy).

*Epistemo-analysis, an extension of Critical Philosophy*
For Critical Philosophy, concepts are operations acting on the world on experience. "Epistemo-analysis" is an extension of Critical Philosophy which unveils and analyses the emotional roots of concepts. This neologism, copied from "psycho-analysis" was introduced recently[2], but the notion existed long before, for instance in Bion's *Theory of Thinking*[3]. Briefly speaking, it makes use of two notions: "family romance" and "object-relation"[4]. Family romance is a way to construct abstract notions like "the past" upon the phantasized subject's own past[5]. Object-relation is a subtle and complex notion (deriving from the Freudian notion of drive) by which the subject at the same time is embedded in a relation with his objects (more exactly "proto-objects") of desire and is detaching himself from this "embeddedness" so that the proto-objects become external objects of desire. In this conception of object-relation, the object is a construction. It must be noticed that the object-relation is logically different from a relation *with* an object. In the object-relation, the relation is *in* the object. Therefore the relationship between the relation and the object is different from the subject/predicate structure of any grammatically correct statement in natural language (analyzed as soon as 1662 in the famous *Logic or Art of Thinking* by Arnauld and Nicolle). That is why it is so difficult to explain the object-relation in natural language whose structure is not adapted to what it is about.

The primary root of embeddedness is affection and objects are "good" objects, or objects of love. This is the unconscious root of living objects and of Life. Empirically it happens that the observable behaviour of these living objects (constructed from object-relation) is correlated with another type of experience, namely physico-chemical experiments of modern biology (like standard organic chemistry). Life as viewed by biology is then an intellectual construction based on physical concepts. Exobiology then tries to find similar observations outside the Earth. But there is no guarantee that we can have emotional relations with these observations.

---

2  See Schneider 2002 and Schneider 2006
3  See also Bion 1962b
4  See Laplanche and Pontalis 1974
5  See its application to Cosmology in Schneider 2006

**"Definition" of exo-Life**

Here we apply the constructive approach to exo-Life. But let us before discuss what is expected from a definition.

*What is a ``definition?*

The word ``definition" is the subject of a very wide literature in philosophy, impossible to summarize in a few paragraphs. It started with Aristotle, followed in the Middle-Age by Nominalism and more recently by different Schools of Logic. In his Posterior Analytics Aristotle discusses the definition as designating the collection of attributes (clearly characterized according to the method of "division") of something (II, Section 3 [Theory of Definition], Chapter XIII). For different adepts of Nominalism (staring with Roscelin of Compiègne, followed by Thomas Aquinus, Pierre Abélard and others) a definition is a name creating a category without seeking for an essence. In contemporary literature of natural sciences a definition essentially refers to two different situations. First, it means an arbitrary convention, like for instance the neologism "pulsar". On the other hand it also often designates an attempt to clarify the content of a pre-existing word for which we have some spontaneous preconceptions, whatever their grounds, and to catch an (illusory) "essence" of what is defined. It is then made use of pre-existing plain language words which carry an a priori pre-scientific content (which can be revealed by Epistemo-Analysis) likely to introduce some confusion in the reader's mind. In a recent attempt, Rosch (1973) tries to put, thanks to to notion of prototype, definitions in full light, even when they are vague. But this approach ignores the empirical fact that words (when they are not pure conventions), and their unconscious (and therefore somehow obscure) content revealed by Epistemo-Analysis, pre-exist any definition.

Modern language theory has pointed out the performative nature of words. They do not really designate pre-existing things, they do create in a first step what they designate as exterior and pre-existing to them in a second step.

Since definition constructs what it defines, there is no absolute definition, only a definition depending on the procedure by which it constructs the *definiendum*. In this sense there is an essential relativity of definitions.

In the remaining part of this paper we will deal with two definitions of Life: a definition based on object-relation and a definition based on standard laboratory Bio-Chemistry (and more generally on Physics).

*Life as a construction and its arbitrariness*

As seen above, Life is not an objective attribute, it is always a construction, based on object-relation in the common-sense meaning of the word Life, or on physico-chemical concepts like in Biology. Therefore, Life, as seen by Astrobiology, is not Life in the object-relation sense. Moreover, Life in the object-relation sense, *i.e.* as an attribute of (unconsciously) emotional relationships, cannot be constructed from purely physico-chemical concepts. Astrobiologists, as physico-cemists working on celestial observations, thus make an improper use of the word Life which inevitably carries the emotional content of object-relation involved in the primitive sense of the word. By doing so, they are fooling the reader[6]. A pertinent analogy is given by the question "When does the human embryo become a human being?" or "When do pre-hominids become humans?". The time at which this transition happen is, unavoidably, an arbitrary choice. To shed a different light on this issue, we note a similarity with Quantum Physics. In Quantum Theory, observables, (represented by linear operators on a vector space) cannot be built from the state vector representing the structure of the measurement apparatus. They are *sui-generis* as pointed out by Ulfbeck and Bohr (2001).

*Organic Life*

There is no essence of Life, even organic. Life, that is the claim that such or such observations reveal that it originates from a living is an arbitrary construction. Experience only consists, like in the object-relation, in relations with objects (constructed out of observations) which we declare (and want to believe) that they are living. Astrobiologists want to declare as living objects which are sufficiently complex and whose complexity is stable and self-regenerated. But such properties also exist for objects recycling matter such as stars, which are not conceived as living. They do just show an amplification of local entropy fluctuations toward less entropy. Therefore objects declared as living in the astrobiological sense of self-organized structures, are not necessarily living in the object-relation (*i.e.* emotional) sense. There is an analogy here with light. When a community of speakers watches a strawberry, it says "it is red". When physicists make a spectral analysis of it, they find a wavelength around 675 nm and there is always a correlation between the plain language word "red" and that wavelength. From this correlation one can identify "red" and 675 nm. But there is no colour associated with wavelengths larger than ~750 nm and smaller than ~ 400 nm. Similarly, there may be not Life in the object-relation sense associated with complex structures (very) different from our terrestrial organisms.

---

6  Like cosmologists who fool the reader by calling « time » the parameter $t$ in Astrophysics (Schneider 2006).

*Intelligent Life.*
For "intelligent" life, we face in addition the paradox that we try to define alien, *i.e.* non-human intelligence in terms of human concepts. It is a kind a paradox like the Zeno paradox: how to analyze motion in terms of static terms, namely a series of static positions. In motion there must be something beyond static positions. It it is the same with extraterrestrial intelligence: human intelligence is a kind of prison which we have to escape. This situation is experienced in SETI in which astrobiologists plan to interpret SETI signals with human concepts. The only hope is to find in ourselves resources beyond standard intelligence, like (psycho-analytic) unconsciousness is beyond consciousness.

**Operational conclusion.**
It is comprehensible that astrobiologists start with some prejudice about exo-life as guidelines for their observations, just because being space-based these observations are very expensive. But at the same time we should keep our minds open and possibly make as much and diverse observations as possible and select from them those with which we can have interesting relations. Like in bio-ethics in which the choice that the embryo is human or not is arbitrary, the claim that such or such observations come from living beings will be arbitrary. Perhaps will we need some day exo-bio-ethical committees, similar to present bio-ethical committees..